\documentclass[fleqn,10pt,onecolumn,preprintnumbers,amsmath,amssymb]{wlscirep}
\usepackage{lipsum}
\usepackage{txfonts}
\usepackage{amsmath}
\usepackage{float}
\usepackage{graphicx}% Include figure files
\usepackage{dcolumn}% Align table columns on decimal point
\usepackage{bm}% bold math
\usepackage{amssymb}
\usepackage{mathrsfs}
\usepackage{amsmath}
\usepackage{bm}
\usepackage{subfigure}
\usepackage{color}
\usepackage{cases}
\title{Probing the Possibilities of Ergodicity in the 1D Spin-1/2 XY Chain with Quench Dynamics}
\author{Hadi Cheraghi}
\author{Saeed Mahdavifar}
\affil{ Department of Physics, University of Guilan, 41335-1914, Rasht, Iran}
\affil[*]{Correspondence and requests for materials should be addressed to H. Cheraghi. (email: \textcolor{blue}{h.cheraghi1986@gmail.com})}

\begin{abstract}
Ergodicity sits at the heart of the connection between statistical mechanics and dynamics of a physical system. By fixing the initial state of the system into the ground state of the Hamiltonian at zero temperature and tuning a control parameter, we consider the occurrence of the ergodicity  with quench dynamics in the one-dimensional (1D) spin-1/2 XY model in a transverse magnetic field. The ground-state phase diagram consists of two ferromagnetic  and paramagnetic  phases. It is known the magnetization in this spin system is non-ergodic.  We set up two different experiments as we call them single and double quenches and test the dynamics of the magnetization along the $Z$-axis  and the spin-spin correlation function along the $X$-axis   which are the order parameters of the  zero-temperature  phases .  Our exact results reveal that for single quenches at zero-temperature, the ergodicity depends on the initial state and the order parameter. In single quenches for a given order parameter, ergodicity will be observed with an ergodic-region for quenches  from another phase, non-correspond to the phase of the order parameter, into itself. In addition, a quench from a  ground-state  phase point corresponding to the order parameter into or very close to the quantum critical point, $h_c=1.0$, discloses an ergodic behavior. Otherwise,  for all other single quenches, the system behaves non-ergodic. Interestingly on the other setup, a double quench on a cyclic path, ergodicity is completely broken for starting from the phase corresponding to the order parameter. Otherwise, it depends on the first quenched point, and the quench time $T$ when the model spent before a second quench in the way back which gives an ability to controlling the ergodicity in the system. Therefore, and contrary to expectations, in the mentioned model  the ergodicity can be observed with probing quench dynamics  at zero-temperature. Our results provide further insight into the zero-temperature dynamical behavior of quantum systems and their connections to the ergodicity phenomenon.
\end{abstract}
\begin{document}
\flushbottom
\maketitle
\thispagestyle{empty}

%%%%%%%%%%%%%%%%%%%%%%%%%%%%%%%%%
%%%%%%%%%%%%%%%%%%%%%%%%%%%%%%%%%
%%%%%%%%%%%%%%%%%%%%%%%%%%%%%%%%%
\section*{Background}
One of the most controversial topics is how the statistical mechanics behavior could emerge in quantum-mechanical systems evolving under unitary dynamics{\color{blue}\cite{1,2,3,4,5,6,7,8,9,10,11, 12}}. Historically, von Neumann was the first one that worked on the topic.  Instead of physical state (or wave function) of the system, he focused on macroscopic observables and introduced the quantum ergodic theorem.  The quantum ergodic theorem says every initial wave function from a microcanonical energy shell evolves so that for most times, in the long run, the joint probability distribution of commuting macroscopic observables obtained from the unitarily time-evolved wave function is close to the microcanonical distribution of commuting observables.

Study of quantum ergodicity in spin systems has been of interest for a long time.
In 1970, for the first time, Barouch and coworkers{\color{blue}\cite{13}} studied the dynamics of the magnetization of the anisotropic spin-1/2 XY chain.  In fact they used a single quench at finite temperature where their initial and final states were thermal states. In addition, they did not probe all quenches. By a quench from the paramagnetic  phase into itself they showed  that the equilibrium is not reached at the final evolutionary time and then the magnetization is a non-ergodic observable. This non-ergodic behavior was later confirmed for the entanglement between the nearest neighbor pair spins of the evolved states{\color{blue}\cite{14}}. In addition to the 1D XY model, the non-ergodicity has been also studied in quantum chaos{\color{blue}\cite{15}},  1D XXZ model to show ergodicity breaking that can create a many-body localization{\color{blue}\cite{16}} and its extended{\color{blue}\cite{17}}, 1D system of spinless and interacting fermions with a disordered potential{\color{blue}\cite{18}}, the anisotropic Dicke model{\color{blue}\cite{ 19}}, and in a small quantum system consisting of three superconducting qubits by measuring the evolution of the entanglement entropy{\color{blue}\cite{20}}.

What connects a physical system to the real world is the unitary time evolution of it. A typical scenario in this context is the quantum quench i.e., driven out-of-equilibrium of a system by abruptly changing a control parameter{\color{blue}\cite{21,22}} where the behavior of the system basically is not susceptible to general principles of equilibrium system{\color{blue}\cite{23,24}}.
The  issue is exciting when a quench is on{\color{blue}\cite{25,26}} or crossed from critical points{\color{blue}\cite{27,28}} where the system undergoes a non-analytic change of its properties. There are some approaches to understanding quench dynamics in many-body systems such as the Kibble-Zurek mechanism{\color{blue}\cite{29}} and measurement quench{\color{blue}\cite{30,31}}.

Here, we are going to study quantum ergodicity  at zero-temperature with the use of quench dynamics of the two  quantities, the magnetization along the $Z$-axis,  and the spin-spin correlation function along the $X$-axis which reveals the magnetization along the $X$-axis. 
We apply two conditions throughout our study: 
 (i) the initial ground state of the Hamiltonian will be  chosen as the initial state of the system (ii) the system will be examined at zero temperature.
 It should note that recently using these conditions, a new approach of quench dynamics was introduced known as the dynamical quantum phase transition{\color{blue}\cite{32}} that has been investigated by using different analytical and numerical techniques{\color{blue}\cite{33,34,35,36,37,38,39,40,41}}, and experimental point of view{\color{blue}\cite{42,43,44}}.
Anyway, using these conditions, we consider 1D spin-1/2 XY model in a transverse field which its ground state phase diagram includes two phases, ferromagnetic (FM)  (with the non-zero value of the magnetization along the $X$-axis) and paramagnetic (PM) (with the almost saturated value of the magnetization along the $Z$-axis). 
Single and double quenches are considered. The long-time run of the  dynamical quantities are compared with their zero-temperature equilibrium values.  
 Our exact results  demonstrate the possibility of the occurrence of ergodicity in the mentioned model.
 We find that, in a single quench, the ergodicity depends on the starting point and  its ordering at zero temperature. For a given order parameter the ergodicity can be observed in two situations, when a quench is done: (i) from another phase, non-corresponding to the phase of the order parameter, into itself which leads to emerge of  an ergodic-region, and (ii) from the phase corresponding to the order parameter into or near to the quantum critical point, $h_{f_1}=h_c$. On the other hand, for all other single quenches, non-ergodic behaviors arise in the system. 
We also discuss the essential effect of the excited states on the creation of the ergodic behavior in the system.
Different behaviors are found for cyclic quenches. In a cyclic quench, the system starts from point $i$ and goes to point $f_1$ by spending a time $T$. After timing spend, the system is quenched into the starting point $f_2=i$.  Based on our results, ergodicity  at zero-temperature will be broken for starting from the phase which is corresponding to the order parameter.
In contrast, for starting from an initial state in a phase irrelevant  to the order parameter, both ergodicity and non-ergodicity can appear depending on the first queched point $f_1$  and the timing spend in this phase point. This outcome gives surprising flexibility in controlling ergodicity by performing double quenches. Consequently, this paper highlights the physical conditions under which the dynamical quantum system may disclose ergodic behaviors at zero-temperature.

%%%%%%%%%%%%%%%%%%%%%%%%%%%%%%%%%
%%%%%%%%%%%%%%%%%%%%%%%%%%%%%%%%%
%%%%%%%%%%%%%%%%%%%%%%%%%%%%%%%%%

\section*{The model}

The Hamiltonian of the 1D spin-1/2 XY model in the presence of a transverse field is given by
\begin{eqnarray}\label{eq1}
{\cal H} = -\frac{1}{2}\sum\limits_{n = 1}^N {\left[ {(1 + \delta )\sigma _n^x\sigma _{n + 1}^x + (1 - \delta )\sigma _n^y\sigma _{n + 1}^y} \right]} - h\sum\limits_{n = 1}^N {\sigma _n^z},
\end{eqnarray}
where $\sigma_n$ is the Pauli spin operator on the $n$-th site,  $\delta$ and $h$ are the anisotropy parameter and the homogeneous external transverse magnetic field, respectively.  $N$ is the system size (or number of spins). The system is considered in the thermodynamic limit, $N \to \infty$. The  quantum phase  transition from the ferromagnetically ordered phase to the paramagnetic phase driven by the transverse field $h$ is called the Ising transition. On the other hand, the  quantum phase transition between two ferromagnetically ordered phases, with magnetic ordering in the  $X$-direction and the Y-direction, respectively, driven by the anisotropy parameter  $\delta$, is called the anisotropic transition. In fact, in the absence of the transverse field, the ground state of the system is in the Luttinger liquid phase at $\delta=0$. Ferromagnetic ordered phase is found in the presence of anisotropy, $0<\delta \leq 1$. The quantum phase transition into the paramagnetic phase is happened at the  quantum critical point $h_c(\delta)= 1${\color{blue}\cite{45,46}}.

This Hamiltonian conserves the parity of the particle number and acts differently on the even (Neveu-Schwarz) and odd (Ramond) subspaces. In the fermionic Fock space, the Hamiltonian in the two subspaces is formally the same if one imposes antiperiodic boundary condition for the even and periodic boundary condition for the odd subspace that in wave-number space these boundary conditions translate to different quantization as momentum quantization  in half-integer and in integer multiples of $\frac{2\pi}{N}$ respectively.
In the thermodynamic limit the ground states of the odd and even subspaces become degenerate and one recovers the two fully polarized ferromagnetic ground states.

\begin{figure}[t]
\centerline{
\includegraphics[width=0.92\linewidth]{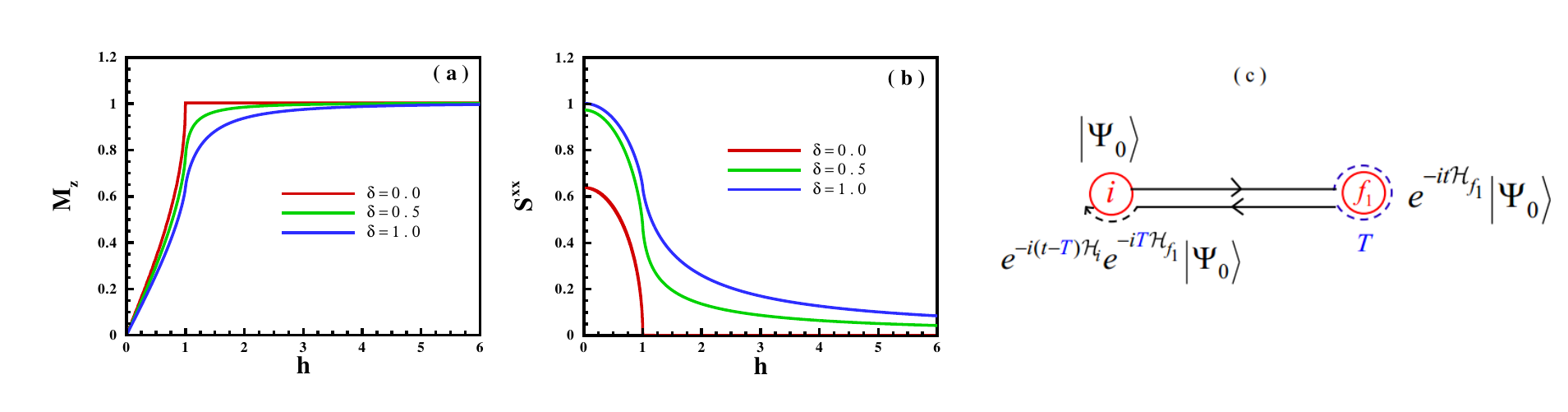}}
\caption{(color online). (a) The magnetization along the Z-axis,  and (b)  spin-spin correlation function along the X -axis versus the transverse magnetic field  at zero temperature at $t=0$. Different values of the anisotropy parameter considered as $\delta=0.0$ (the XX model), $\delta=0.5$ (the XY model), $\delta=1.0$ (the Ising model). (c) The schematic diagram of a cyclic quench.   }
\label{Fig1}
\end{figure}

The Hamiltonian  is  integrable and can be mapped to a system of free fermions and therefore be solved exactly. By applying the Jordan-Wigner transformation{\color{blue}\cite{47}},
the Hamiltonain converts from spin operators into spinless fermionic operators as
\begin{eqnarray}\label{eq2}
{\cal H} =  - \sum\limits_{n = 1}^N {\left[ {\delta \left( {a_n^\dag a_{n + 1}^\dag  + {a_{n + 1}}{a_n}} \right) + \left( {a_n^\dag {a_{n + 1}} + a_{n + 1}^\dag {a_n}} \right)} \right.} + \left. {h\left( {2a_n^\dag {a_n} - 1} \right)} \right],
\end{eqnarray}
where $a_n$ is fermionic operator. Now, performing a Fourier transformation as ${a_n} = \frac{1}{{\sqrt N }}\sum\limits_k {{e^{ - ikn}}}~ {a_k}$, and also Bogoliobov transformation as ${a_k} = \cos ({\theta _k})~{\alpha _k} + i\sin ({\theta _k})~\alpha _{ - k}^{\dag}$ lead to the quasi-particle diagonalized  Hamiltonian  as
\begin{equation}
{\cal H} = \sum\limits_k {{\varepsilon }_k\left( {\alpha _k^\dag {\alpha _k} - \frac{1}{2}} \right)}
\end{equation}
where the energy spectrum is $\varepsilon_k = \sqrt{{\cal A}_k ^2+{\cal B}_k ^2}$ with
\begin{eqnarray}
{\cal A}_k=-2[\cos(k)+h] ~~~~~~~~;~~~~~~~~{\cal B}_k= 2 \delta \sin(k),
\end{eqnarray}
and  $\tan (2{\theta _k}) =- \frac{{{{\cal B}}_k }}{{{{\cal A}}_k }}$. The zero-temperature expectation  values of the magnetization along the $Z$-axis  and the spin-spin correlation function along the $X$-axis at $t=0$ as a function of the transverse field are shown in Fig.~{\color{blue}\ref{Fig1}} (a $\&$ b). It should be noted that these two quantities are defined as

\begin{eqnarray}
M_z=\frac{1}{N} \sum\limits_{n = 1}^N\langle  \sigma _n^z \rangle= \frac{1}{N}  \langle \sigma_{tot}^z \rangle ~~~~~~~~;~~~~~~~~
S^{xx} = \frac{1}{N}\sum\limits_{n = 1}^N {\left\langle {\sigma_n^x\sigma_{n + 1}^x} \right\rangle ,} 
\end{eqnarray}
where the brackets symbolize the expectation value  on the ground state of the system in the thermodynamic limit. As is seen in Fig.~{\color{blue}\ref{Fig1}} (a $\&$ b), the quantum phase transition occurs in the critical transverse field $h_c=1$. Only at $\delta=0$, the magnetization will be saturated (Fig.~{\color{blue}\ref{Fig1}} (a)), and the spin-spin correlation function will be vanished (Fig.~{\color{blue}\ref{Fig1}} (b))  at the critical field.

%%%%%%%%%%%%%%%%%%%%%%%%%%%%%%%
%%%%%%%%%%%%%%%%%%%%%%%%%%%%%%%%

%%%%%%%%%%%%%%%%%%%%%%%%%%%%%%%%%
%%%%%%%%%%%%%%%%%%%%%%%%%%%%%%%%%
%%%%%%%%%%%%%%%%%%%%%%%%%%%%%%%%%
\section*{Setup and  ergodicity}

It is known that, ergodicity in the systems depends on the path where systems evolve with time. Here, we setup two strategies for studying ergodicity in the system. First, in the view of single quench as the tendency of dynamics of the system to match itself to the zero-temperature value of the final quenched point. Accordingly, the system needs to be investigated in a long-time run where it goes or fluctuates around a stable situation.  Second, we apply the idea of double quenches to consider ergodicity in the system.
 For this case, we do:\\
\\
(1) A quench from an initial state in  the phase $A$ to a final phase point in the phase $B$ as $i$  to $f_1$,  \\
\\
(2) A quench from an initial state in  the phase $B$ to a final phase point in the phase $A$ as $f_1$ to $f_2=i$. \\
\\
It means, the starting and end points are the same way (a cyclic quench). The setup is schematically represented in Fig.~{\color{blue}\ref{Fig1}} (c).
 In this way, our instrumentations are  the magnetization along the Z -axis and the spin-spin correlation
function along the X -axis. The time-dependent Hamiltonian $H(t)$ that models a double quantum quench is
\begin{eqnarray}
{\cal H}(t) = \left\{ {\begin{array}{*{20}{c}}
{\begin{array}{*{20}{c}}
{{{\cal H}_i}}\\
{{{\cal H}_{{f_1}}}}\\
{{{\cal H}_{{f_2}}}}
\end{array}}&{\begin{array}{*{20}{c}}
{t \le 0}\\
{0 \le t \le T}\\
{t \ge T}
\end{array}}
\end{array}} \right.
\end{eqnarray}
with $\left| {{\Psi _i}} \right\rangle$ and $\left| {{\Psi _{{f_1}}}(T)} \right\rangle  = {e^{ - iT{{\cal H}_{{f_1}}}}}\left| {{\Psi _i}} \right\rangle $ that are the initial state at $t=0$ and quenched state at $t=T$, respectively.
We investigate our problem with two conditions: \\
\\
(i) Fixing the initial state of the system into the ground state of the Hamiltonian.  \\
\\
(ii) Considering the system at zero temperature. \\
\\

It should note it was demonstrated that by the use of a double quench one can control the dynamical quantum phase transitions in the 1D spin-1/2 ITF model{\color{blue}\cite{48}}. This controlling can be done simply by tuning the time between the first and the second quench. As a result, the system can exhibit all four combinations of absence or presence of non-analyticities before and after the second quench in the  rate function of the return probability, respectively. In the following, as will be seen, the same situation for controlling ergodicity in the system will be achievable by controlling the quench time $T$ in the dynamical behavior of the order parameters.

%%%%%%%%%%%%%%%%%%%%%%%%%%%%%%%
\begin{figure*}[t]
\centerline{
\includegraphics[width=0.85\linewidth]{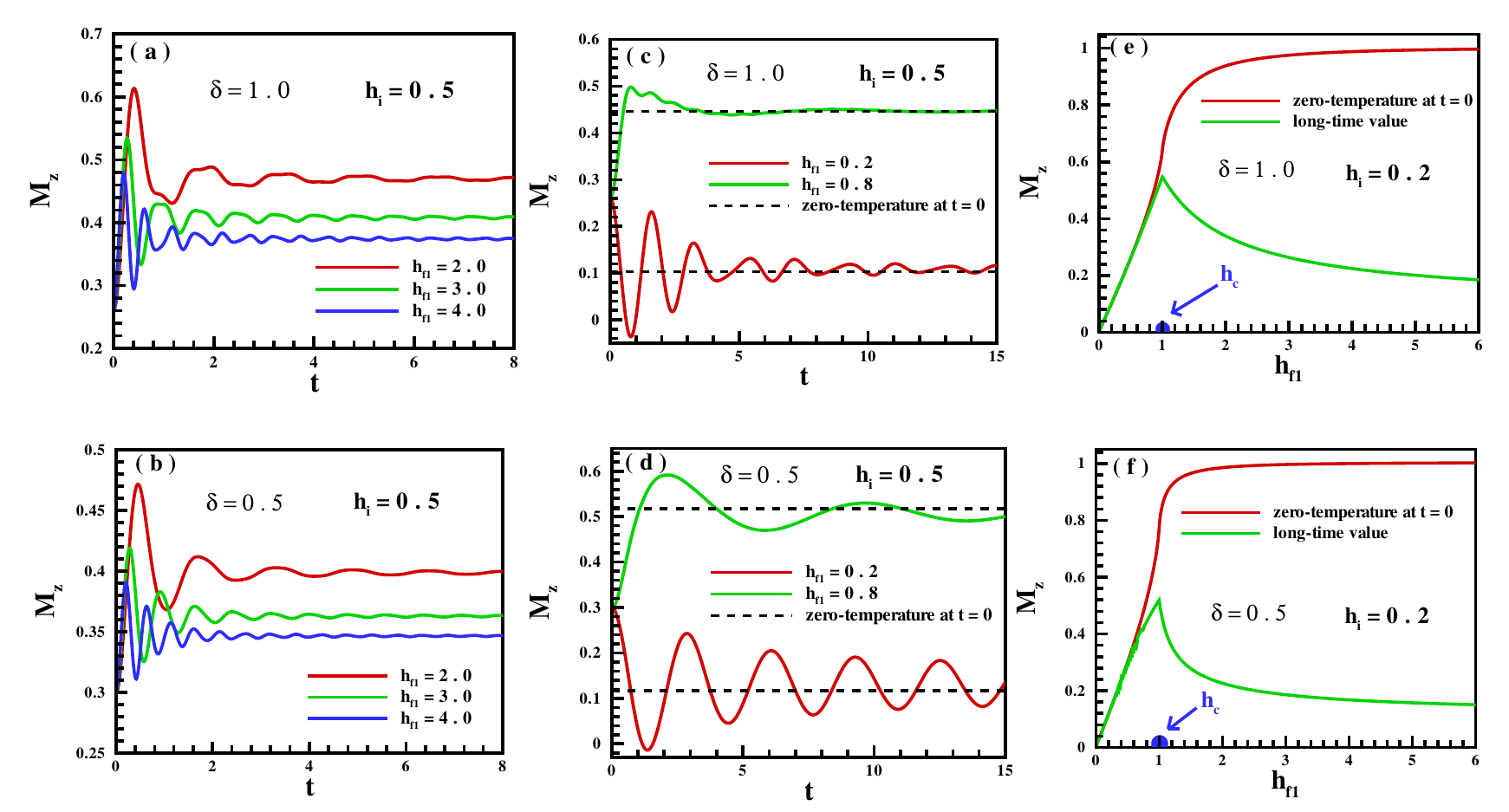}}
\caption{(color online). The time evolution of the magnetization for single quenches as the FM phase into (a $\&$ b)  the PM phase, and (c $\&$ d)  the same FM phase for $\delta=1.0$ and $\delta=0.5$ respectively.  The magnetization for  the long-time run dynamics versus the magnetic field for single quenches  started from the FM region into all points of the quantum states for (e) $\delta=1.0$, and (f) $\delta=0.5$. }
\label{Fig2}
\end{figure*}
%%%%%%%%%%%%%%%%%%%%%%%%%

%%%%%%%%%%%%%%%%%%%%%%%%%%%%%%%%%
%%%%%%%%%%%%%%%%%%%%%%%%%%%%%%%%%
%%%%%%%%%%%%%%%%%%%%%%%%%%%%%%%%%

\section*{Ergodicity in a single quench}

Based on the quantum ergodic theorem,  for an arbitrary initial state, the expectation value of the magnetization and the spin-spin correlation function  will ultimately evolve in time to its value predicted by the theory of ensemble, and thereafter will exhibit only small fluctuations around equilibrium value.
To consider the time evolution of a closed quantum system, one can use of  quench dynamics,  putting the system in an equilibrium state described with the Hamiltonian $\mathcal{H}_i = \sum\limits_k {{\mathcal{H}_k^i}} $ and the initial state  $\left| {{\Psi _0}} \right\rangle $, afterward,  suddenly changing the control parameters from their initial values to their final values. Final Hamiltonian and its time-evolved state will be as  $\mathcal{H}_{f_1} = \sum\limits_k {{\mathcal{H}_k^{f_1}}} $ and $\left| {\Psi (t)} \right\rangle  = {e^{ - i\mathcal{H}_{f_1}t}}\left| {{\Psi _0}} \right\rangle $, respectively. We choose the ground state of the initial Hamiltonian as the initial state, and study the system  at zero temperature.

In the following, we  focus on the dynamics of the magnetization along the transverse field  and the spin-spin correlation function along the $X$-axis. The  magnetization along the $Z$-axis is the order parameter of the PM phase. On the other side, the spin-spin correlation function along the $X$-axis reveals the ordering of the FM phase which is analogous  to the magnetization along the $X$-axis.
For a single quench where $t \le T$,  we have obtained the time-dependent of these quantities  in the thermodynamic limit as
\begin{eqnarray}
M^z(t)&=&   -\frac{2}{N}\sum \limits_{k > 0} \left[  \cos(2\theta _k^{f_1})\cos(2\Phi _1^k) + \sin(2\theta _k^{f_1})\sin (2\Phi _1^{k})\cos (2\varepsilon _k^{f_1} t) \right], \nonumber \\
S^{xx}(t) &=&  -\frac{2}{N}\sum\limits_{k > 0} {\left[ { \cos (2\Phi _1^k)\cos (k + 2\theta _k^{{f_1}}) + \sin (2\Phi _1^k)\sin (k + 2\theta _k^{{f_1}})\cos (2\varepsilon _k^{{f_1}}t)} \right]}, 
\label{eq22}
\end{eqnarray}
where  ${\Phi _1^k} = \theta _k^{f_1}- \theta _k^i$ is the difference between the Bogoliubov angles diagonalizing the pre-quench and post-quench Hamiltonians, respectively.

%%%%%%%%%%%%%%%%%%%%%%%%%%%%%%%%%%
\begin{figure}[t]
\centering{
\includegraphics[width=0.555\linewidth]{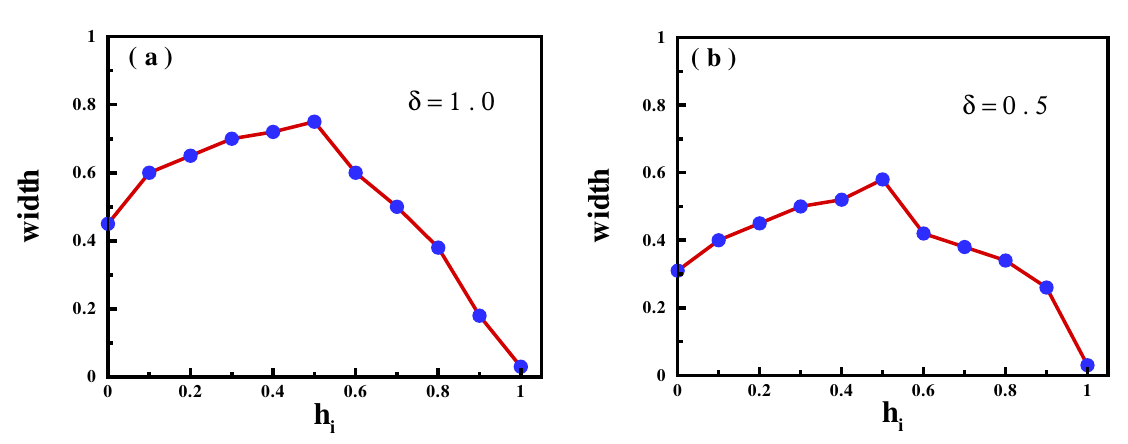}}
\caption{(Color online) Width of the ergodic-region for magnetization for quenches starting from the initial states in the FM phase for (a) $\delta=1.0$, and (b) $\delta=0.5$.}
\label{Fig3}
\end{figure}
%%%%%%%%%%%%%%%%%%%%%%%%%%%%%%%%

%%%%%%%%%%%%%%%%%%%%%%%%%%%%%%%%
\begin{figure}[t]
\centering{
\includegraphics[width=0.57\linewidth]{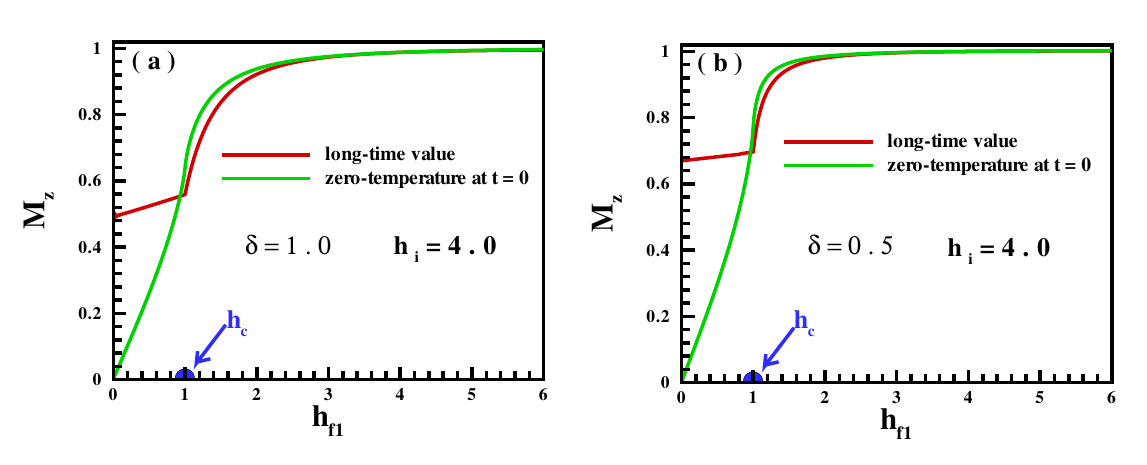}}
\caption{(Color online) The magnetization for  the long-time run dynamics versus the magnetic field for single quenches  started from the PM region into all points of the quantum states for (a) $\delta=1.0$, and (b) $\delta=0.5$. }
\label{Fig4}
\end{figure}
%%%%%%%%%%%%%%%%%%%%%%%%%%%%%%%%%

In the first step, we have focused on the dynamical behavior of the magnetization.
We have considered  a quench from  an initial state in the FM phase. Results are illustrated  in Fig.~{\color{blue}\ref{Fig2}} for different values of $\delta=1.0,~0.5$ and $h_i=0.5$. We know that the zero-temperature  state of the system is in the FM phase at $h_i=0.5$ and at the value of field $h>h_c=1.0$ goes into the PM phase where all spins are aligned along the transverse magnetic field.  As is apparent from Fig.~{\color{blue}\ref{Fig2}} (a $\&$ b), for a single quench from  the mentioned initial state in  the FM phase into the PM region, the value of the magnetization for most times in the long run is very far from the  zero-temperature saturated phase. Thus the spin-1/2 XY chain for the mentioned initial state behaves as a non-ergodic system in quenching from the FM into the PM regions. For a quench from the FM into the same  FM region, we also have calculated the magnetization and results are presented in Fig.~{\color{blue}\ref{Fig2}} (c $\&$ d) for $h_i=0.5$. As is obvious, the system shows ergodic behavior for selected values of the transverse magnetic field. In fact, the magnetization ultimately evolves in time to its value predicted by ground state of quenched Hamiltonian, and thereafter will exhibit small fluctuations around zero-temperature value.  We did the same quenches starting from an initial state in the FM region into all phase points. Results for the long-time run dynamics of the  magnetization are displayed in  Fig.~{\color{blue}\ref{Fig2}} (e $\&$ f) for $\delta=1.0,~0.5$ and $h_i=0.2$. The phenomenon of broken ergodicity is manifestly seen in this figure. In fact, by starting from a selected initial state in the FM phase and quenching into the same  FM region, the system does not break the ergodicity in a wide region. Approaching the quenched point to  the quantum critical region, deviation from the ergodic behavior starts to reveal. This phenomenon is obviously shown for quenching into a phase point in the PM region. 

We have calculated the width of the ergodic-region  of magnetization for different initial points selected in the FM region. The results are demonstrated in Fig.~{\color{blue}\ref{Fig3}} (a $\&$ b).  As is seen, the width of the ergodic-region displays a non-monotonic behavior and  will be maximized at $h_i=\frac{h_c}{2}=0.5$. This may be attributed to the fact that at this initial point, the system has a large coherence and behaves as a quantum memory which is capable of storing the quantum information{\color{blue}\cite{49, 50, 51}}. 
Furthermore, approaching the quantum critical region, the width of the ergodic region decreases and will be minimized at the quantum critical point, $h_i=h_c=1$. In other words,  its width at the quantum critical point is approximately zero that indicates at quenching from $h_c$ into both the FM and the PM phase, system behaves non-ergodic. 

In the second step, a quench from an initial state in the PM phase into different regions is considered. Results on the long-time run of magnetization are indicated  in Fig.~{\color{blue}\ref{Fig4}} for different values of $\delta=1.0,~0.5$ and $h_i=4.0$. It is explicitly  shown that, only in a quench into a phase point very close to the quantum critical point,  the system unveils  an ergodic behavior. We have to mention that there is not any overlap between curves in the high magnetic region. One conclusion that can be drawn from the discussion up to now is that, the magnetization along  the transverse
magnetic field can be exhibited an ergodic behavior if a quench is done from a region  where the magnetization:\\
\\
(1) is not the order parameter into the same region,\\
\\
(2) is  the order parameter into the quantum critical point $h_{f_1}=h_c$.
\\

 Let us more explain the role of the excited states emerging from the dynamics of the system to create ergodicity. The magnetization along $Z$-axis as a function of $t$ can be rewritten as
\begin{eqnarray}
M^z(t)&=& \frac{1}{N} ~\left[ | C_0 |^2 \left\langle E_0 ^{f_1} | \sigma^{z}_{tot} | E_0 ^{f_1} \right\rangle  + \sum_{n \neq 0} ( C_0^{*} C_n e^{-i(E_n ^{f_1}- E_0 ^{f_1})t} \left\langle E_0 ^{f_1} | \sigma^{z}_{tot} | E_n ^{f_1} \right\rangle + h.~c. )\right. \nonumber \\
&+&\left.   \sum_{n,n'  \neq 0}  C_{n'}^{*} C_n e^{-i(E_n ^{f_1}- E_{n'} ^{f_1})t} \left\langle E_{n'} ^{f_1} | \sigma^{z}_{tot} | E_n ^{f_1} \right\rangle  \right], \nonumber \\  
\label{eq23}
\end{eqnarray}   
where the eigenvalues $\{E_n^{f_1}\}$ and eigenvectors $\{   \left| E_n^{f_1} \right\rangle   \}$ with $n=0,1,2,...,$ correspond to the Hamiltonian  at the middle quenched point which can be obtained from the Schr\"odinger equation, $\mathcal{H}_{f_1}\left| E_n^{f_1} \right\rangle = E_n^{f_1} \left| E_n^{f_1} \right\rangle $. Hence, $E_0^{f_1} $ is the ground state energy of the system at quenched point $f_1$ and the expansion coefficients are $C_n = \left\langle {E_n^{{f_1}}|E_0^i} \right\rangle $.
The system shows ergodic behavior if, at the long-time run, the value of the magnetization becomes equal to  zero-temperature value in the ground state of the quenched point $f_1${\color{blue}\cite{2,52}}

%%%%%%%%%%%%%%%%%%%%%%%%%%%%
\begin{figure}[t]
\centering{
\includegraphics[width=0.57\linewidth]{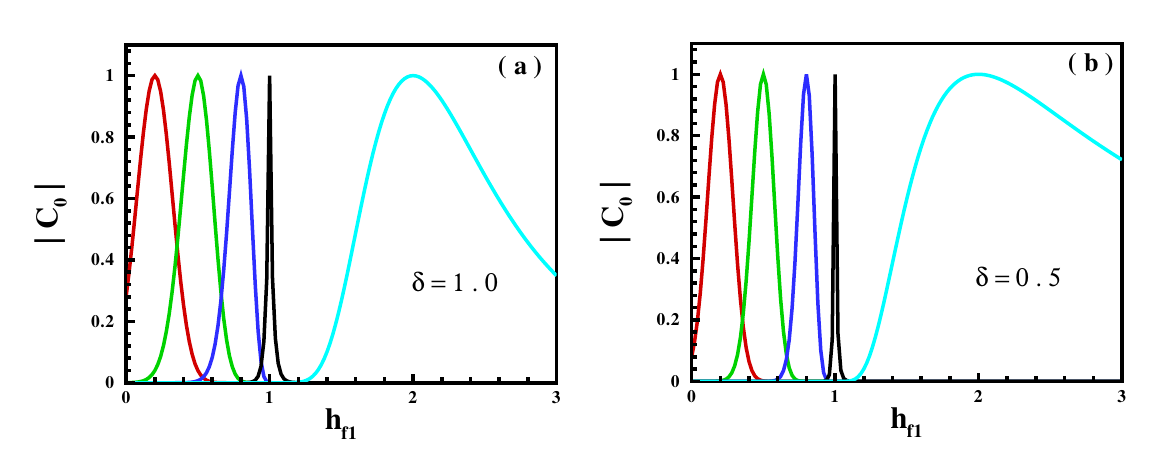}}
\caption{(Color online)  The expansion coefficient $|C_0|$ for quenches from $h_i=0.2,0.5,0.8,1.0,2.0$ (from red to cyan) to $h_{f_1}$ for (a) $\delta=1.0$, and (b) $\delta=0.5$.}
\label{Fig5}
\end{figure}
%%%%%%%%%%%%%%%%%%%%%%%%

\begin{eqnarray}
\lim_{t\longrightarrow \infty}   M^z(t)= \frac{1}{N} \left\langle E_0 ^{f_1} | \sigma^{z}_{tot} | E_0 ^{f_1} \right\rangle . 
\label{eq24}
\end{eqnarray}
In principle, at $t \to \infty$, the  Eq. ({\color{blue}{\ref{eq23}}}) transvers to 
\begin{eqnarray}
{M^z}(t)= \frac{1}{N}|{C_0}{|^2}\left\langle {E_0^{{f_1}}} \right|\sigma _{tot}^z\left| {E_0^{{f_1}}} \right\rangle  +\frac{1}{N} \sum\limits_{n \ne 0}^N |C_n|^2\left\langle E_{n} ^{f_1} | \sigma^{z}_{tot} | E_n ^{f_1} \right\rangle ,
\label{eq25}
\end{eqnarray}
that indicates if  a system behaves as a non-ergodic system and  if $|C_0|=1$, thereby, the  excited states are ineffective in the ergodicity of the system ($|C_n|$  must be zero for all ${n \ne 0}$).
In Fig.~{\color{blue}\ref{Fig5}} the coefficient $|C_0|$  is plotted for $\delta=1.0,~0.5$ that clarity shows $|C_0|=1$ only happens when no quench is done, otherwise, $|C_0|<1$. In the other side, because it should be $\sum\limits_{n = 0}^N {|{C_n}{|^2} = 1}$, consequently, the ergodic behavior arises if at least there is a $|C_n|\ne 0$  for ${n \ne 0}$ in such a way that Eq. ({\color{blue}{\ref{eq25}}}) can satisfy  Eq. ({\color{blue}{\ref{eq24}}}). As a result, it reveals for quenching into the PM phase, it is ever zero the existence of the excited states which lead to the behavior of ergodicity in the system. It also expresses  for quenching into the FM phase, only the excited states in the ergodic-region have the main role in the ergodicity  of the system.

%%%%%%%%%%%%%%%%%%%%%%%%%%%%%
\begin{figure}[t]
\centering{
\includegraphics[width=0.57\linewidth]{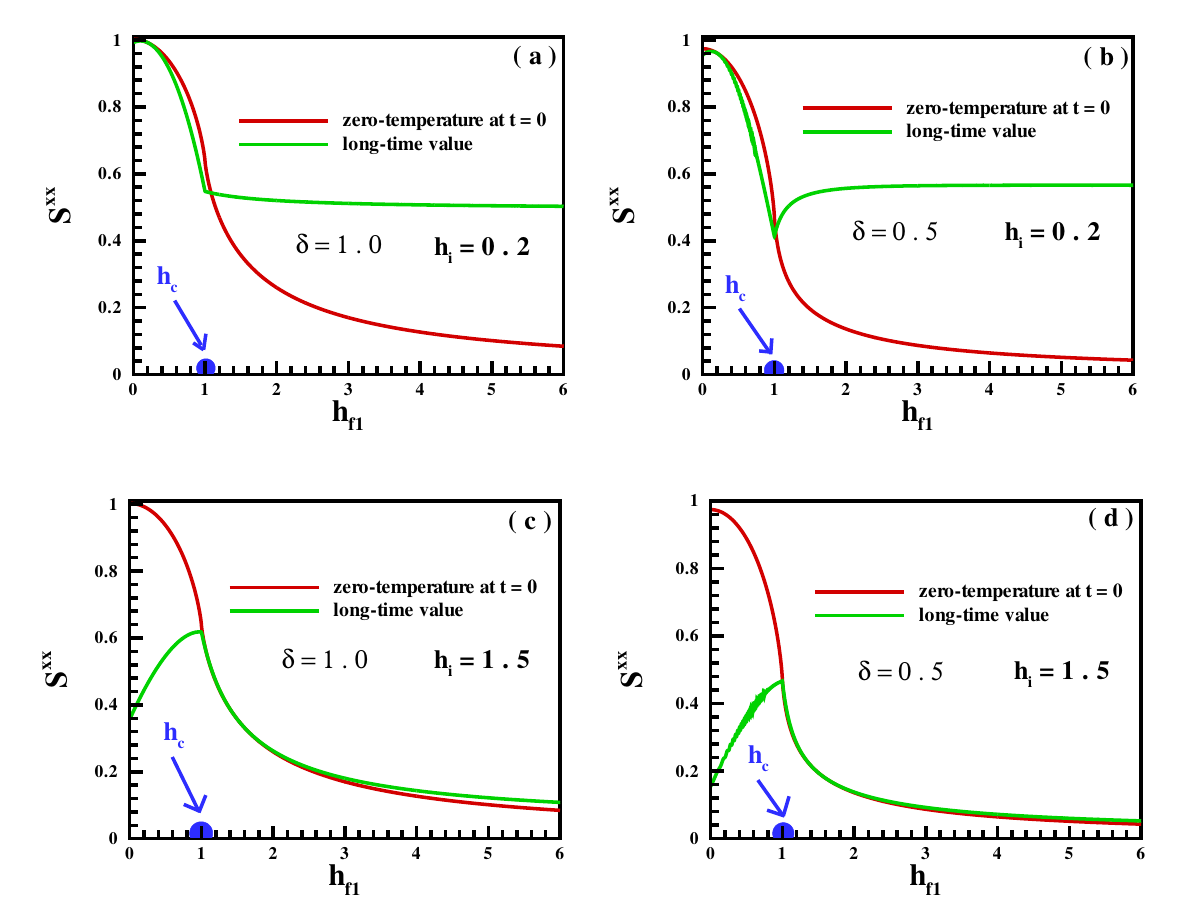}}
\caption{(Color online) The  spin-spin correlation function along the $X$-axis  for  the long-time run dynamics versus the  transverse magnetic field for single quenches  started from (a $\&$ b) the FM region, and (c $\&$ d) the PM region into all points of the quantum states for $\delta=1.0$ and $\delta=0.5$, respectively. }
\label{Fig6}
\end{figure}
%%%%%%%%%%%%%%%%%%%%%%%%%%%%%

Another result that lies in Fig.~{\color{blue}\ref{Fig2}} (e $\&$ d)  is   quenching into the PM phase where for a very large value of the magnetic field it leads to vanishing the long-time run of the magnetization. In this case, since in the PM phase the spins of the system behave as a non-interacting many-particle system, the quenched Hamiltonian can be approximately considered as the only Zeeman interaction, ${\cal H}_{f_1}  \simeq - h_{f_1}\sum_{n} \sigma_{n}^{z}$, with $\left| {E_0^{{f_1}}} \right\rangle  = \left| { \uparrow  \uparrow ... \uparrow } \right\rangle $,
thus,  Eq. ({\color{blue}{\ref{eq25}}}) gives
\begin{eqnarray}
{M^z}(t) = \sum\limits_{n = 0}^N {\left( {1 - \frac{{2n}}{N}} \right)\left( {\begin{array}{*{20}{c}}
N\\
n
\end{array}} \right)|{C_n}{|^2}} 
\label{eq26}
\end{eqnarray}
inasmuch as the probability of the number of $N-n$ spins are up is equal to the number of  $n$ spins are down, $|{C_n}| = |{C_{N - n}}|$, subsequently, the Eq. ({\color{blue}{\ref{eq26}}}) will be zero. From the view point of the spin wave collective excitations, in the ground state of the XY model in the PM region, only, very low-energy spin wave excitations are accompanied (please see Fig.~{\color{blue}\ref{Fig1}} (a)), which show that by quenching into the PM region, a small density of spin wave excitations will be created. On the other hand, the initial state with the saturated FM ordering along the $X$-axis is a superposition of all spin wave excitations with the same probability. Therefore, the system behaves non-ergodic when a quench into the PM phase will be done. 

Now, let us calculate the particular modes in the ergodic-region. Another expression of the Eq. ({\color{blue}{\ref{eq24}}}) is
\begin{eqnarray}
\overline {{M^z}(t)}  = \lim_{T\longrightarrow \infty} \frac{1}{T}\int_0^T {dt{M^z}(t)} 
\end{eqnarray}
Thus, from  Eq. ({\color{blue}{\ref{eq22}}}) for the ergodic-region  one can obtain
\begin{eqnarray}
\cos (2\theta _{{\kappa ^{*}}}^{{f_1}})\left[ {1 - \cos (2\Phi _1^{{\kappa ^{*}}})} \right] = 0,
\label{eq27}
\end{eqnarray}
that it gives $\kappa ^* = m\pi $ with $m = 0, \pm 1, \pm 2,...,$ and $\kappa ^* = \arccos ( - {h_{{f_1}}})$. It should be stressed, comparing   these  modes with  the particular modes which lead to the dynamical quantum phase transitions {\color{blue}\cite{32}}  discloses that they are different. 

In addition to the magnetization along the transverse field, we  have also studied  the dynamical behavior of the spin-spin correlation function along the $X$-axis. Results are exhibited in Fig.~{\color{blue}\ref{Fig6}} for  the long-time run dynamics versus the  transverse magnetic field for single quenches  started from (a $\&$ b) the FM region with $h_i=0.2$, and (c $\&$ d) the PM region with $h_i=1.5$, into all points of the quantum states for $\delta=1.0$ and $\delta=0.5$ respectively. In general, this quantity behaves inversely with respect to the magnetization along the transverse field. Starting from an initial state in the FM region, no ergodic behavior is seen except in a quench into a final state very close to the quantum critical point, $h_c=1.0$. But, when we  put the system initially in the PM phase, an ergodic region is observed for quenching into a final state in the same PM phase.

%%%%%%%%%%%%%%%%%%%%%%%%%%%%%%%%%%
\begin{figure}[t]
\centering{
\includegraphics[width=0.56\linewidth]{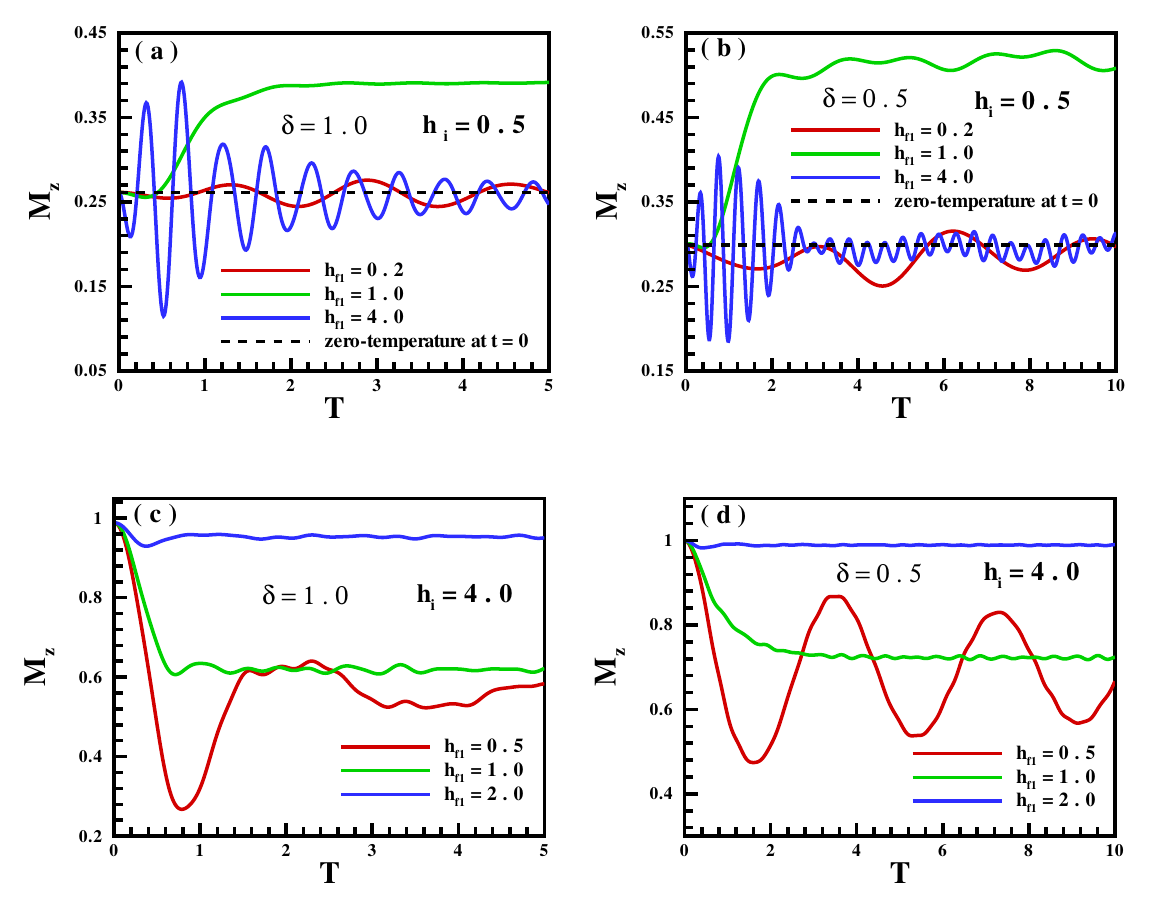}}
\caption{(Color online)   Long-time run  of the magnetization versus the quench time $T$  for cyclic quenches  from (a $\&$ b) FM phase, (c $\&$ d) PM phase, into regions of the FM, PM and the critical point for $\delta=1.0$ and $\delta=0.5$ respectively. }
\label{Fig7}
\end{figure}
%%%%%%%%%%%%%%%%%%%%%%%%%%%%%%%%%%

%%%%%%%%%%%%%%%%%%%%%%%%%%%%%%%%%
%%%%%%%%%%%%%%%%%%%%%%%%%%%%%%%%%
%%%%%%%%%%%%%%%%%%%%%%%%%%%%%%%%%

\section*{Ergodicity in a cyclic quench}

As we have mentioned, a cyclic quench is defined as a linked sequence of quenches where eventually returns the system to its initial starting point.  The magnetization  along the transverse field  and the spin-spin correlation function along the $X$-axis at zero temperature as a function of time in a double quench ($t \ge T$) are obtained as
\begin{eqnarray}
M_z(t) &=&  \frac{4}{N}\sum\limits_{k>0} \left[Q_1^k(t)^2 +Q_2^k(t)^2-\frac{1}{2} \right], \nonumber\\
S^{xx}(t) &=& \frac{4}{N}\sum\limits_{k>0} \left[\cos(k) \left(Q_1^k(t)^2 +Q_2^k(t)^2 \right)+\sin(k) \left(Q_1^k(t)P_1^k(t)-Q_2^k(t)P_2^k(t)  \right) \right], 
\end{eqnarray}
where
\begin{eqnarray}
Q_1^k(t) &= &\left[\sin (\theta _k^{f_2})\cos (\Phi _1^k) - \cos (\theta _k^{f_2})\sin (\Phi _1^k) \right] \cos (\Phi _2^k)\cos \left( (t-T)\varepsilon _k^{f_2} + T\varepsilon _k^{f_1}\right)\nonumber\\
&-&\left[ \cos (\theta _k^{f_2})\cos (\Phi _1^k) + \sin (\theta _k^{f_2})\sin (\Phi _1^k) \right]   \sin (\Phi _2^k)\cos \left( (t-T)\varepsilon _k^{f_2} - T\varepsilon _k^{f_1}\right),
\end{eqnarray} 

\begin{eqnarray}
Q_2^k(t) &= &\left[ \sin (\theta _k^{f_2})\cos (\Phi _1^k) + \cos (\theta _k^{f_2})\sin (\Phi _1^k)\right]  \cos (\Phi _2^k)\sin \left((t-T)\varepsilon _k^{f_2} + T\varepsilon _k^{f_1}\right)   \nonumber\\
&+&\left[ \cos (\theta _k^{f_2})\cos (\Phi _1^k) - \sin (\theta _k^{f_2})\sin (\Phi _1^k)\right]    \sin (\Phi _2^k)\sin \left((t-T)\varepsilon _k^{f_2} - T\varepsilon _k^{f_1}\right),
\end{eqnarray}

\begin{eqnarray}
P_1^k(t) &= & \left[ {\cos (\theta _k^{{f_2}})\cos (\Phi _1^k) +\sin (\theta _k^{{f_2}})\sin (\Phi _1^k)} \right]\cos (\Phi _2^k)\cos \left((t-T)\varepsilon _k^{{f_2}} + T\varepsilon _k^{{f_1}}\right)\nonumber\\
 &+& \left[ {\sin (\theta _k^{{f_2}})\cos (\Phi _1^k) - \cos (\theta _k^{{f_2}})\sin (\Phi _1^k)} \right]\sin (\Phi _2^k)\cos \left((t-T)\varepsilon _k^{{f_2}} - T\varepsilon _k^{{f_1}}\right),
\end{eqnarray} 

\begin{eqnarray}
P_2^k(t) &= & {\left[ {-\cos (\theta _k^{{f_2}})\cos (\Phi _1^k) + \sin (\theta _k^{{f_2}})\sin (\Phi _1^k)} \right]\cos (\Phi _2^k)\sin \left((t-T)\varepsilon _k^{{f_2}} + T\varepsilon _k^{{f_1}}\right)} \nonumber\\
 &+&  {\left[ {\sin (\theta _k^{{f_2}})\cos (\Phi _1^k) + \cos (\theta _k^{{f_2}})\sin (\Phi _1^k)} \right]\sin (\Phi _2^k)\sin \left((t-T)\varepsilon _k^{{f_2}} - T\varepsilon _k^{{f_1}}\right)},
\end{eqnarray}
with $\Phi _2^k={\theta _k^{f_2}}-{\theta _k^{f_1}}$.  
Here, for investigating that in a cyclic quench, the system exposes  ergodic behavior or not, we have considered two different initial situations. Once, we have put the system initially in the FM phase, once again we have considered it in the PM phase and done a linked sequence of quenches ($i \longrightarrow f_1 \longrightarrow i $) where finally the system after passing time $T$ returns to its initial state. 

Long-time run  of the magnetization is indicated in  Fig.~{\color{blue}\ref{Fig7}} for different values of the anisotropy parameter. As is seen in Fig.~{\color{blue}\ref{Fig7}} (a $\&$ b), in a cyclic quench started from FM phase ($h_i=0.5$), if in during double quenches the system goes to the quantum critical point ($h_{f_1}=h_c=1$) and spends a time $T$ (not very short) at the quantum critical point, in returning into the initial point, our system shows the non ergodic behavior. Otherwise, depending on the value of $h_{f_1}$, the system may or may not show ergodic behavior. However, as explicitly illustrated, the ergodicity of the system is oscillatory in such a way that it happens for the special value of $T$. It means by controlling the quench time $T$, one can control the ergodicity in the system. This comes from the fact that the system acts as an
infinite memory of its past state, and gives an opportunity to control whether or not the system turns back to its initial situation after the second quench. In other words, these results are similar to those that obtained to controlling the non-analyticities for the rate function
of the return probability{\color{blue}\cite{48}}.
As another result, the value of $T$, depends on the value of $h_{f_1}$. For example in  Fig.~{\color{blue}\ref{Fig7}} (a $\&$ b),  for the first ergodicity time $T$, we have $T_{h_{f_1}=4.0}<T_{h_{f_1}=0.2}$. 
 On the other note, when the system is initially located in the PM phase, different behavior is observed.  Fig.~{\color{blue}\ref{Fig7}} (c $\&$ d), shows that system behaves non-ergodic in all cyclic quenches independent of the anisotropy parameter, the middle quenched point, and spending time $T$.

%%%%%%%%%%%%%%%%%%%%%%%%%%%%%
\begin{figure}[t]
\centering{
\includegraphics[width=0.57\linewidth]{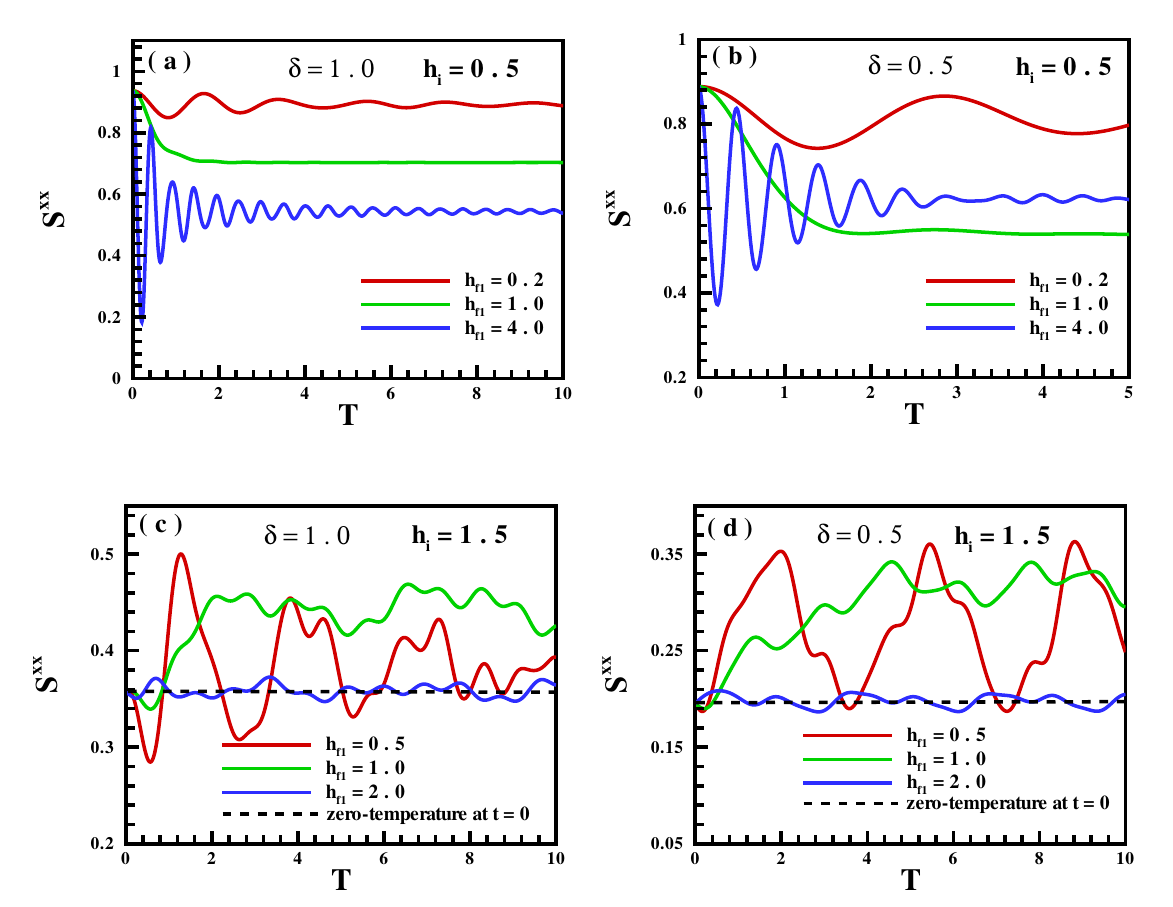}}
\caption{(Color online) The   spin-spin correlation function along the $X$-axis  for  the long-time run dynamics versus the  transverse magnetic field for single quenches  started from (a $\&$ b) the FM region, and (c $\&$ d) the PM region into all points of the quantum states for $\delta=1.0$ and $\delta=0.5$ respectively. }
\label{Fig8}
\end{figure}
%%%%%%%%%%%%%%%%%%%%%%%%%%%%%

We  have also investigated  the dynamics of the spin-spin correlation function along the $X$-axis for cyclic quenches. Fig.~{\color{blue}\ref{Fig8}} displayed the results for  the long-time run dynamics versus the quench time $T$  started from (a $\&$ b) the FM region with $h_i=0.5$, and (c $\&$ d) the PM region with $h_i=1.5$, into $h_{f_1}=0.5,1.0,2.0$ for $\delta=1.0$ and $\delta=0.5$ respectively. As we mentioned, in general, this quantity behaves conversely rather than the magnetization along the transverse field. Furthermore, in one state ergodicity comes out. This is when a quench is started from the PM phase and the first quenched point $h_{f_1}$  will be as  an appropriate phase point where results in the emerge of ergodicity with an ability to control it with quench time $T$. For other quenches, including quench to the critical point (for not very short time T), the system behaves non-ergodic.

%%%%%%%%%%%%%%%%%%%%%%%%%%%%%%%%%%
%%%%%%%%%%%%%%%%%%%%%%%%%%%%%%%%%%
%%%%%%%%%%%%%%%%%%%%%%%%%%%%%%%%%%
\section*{Conclusion}

The phenomenon of the ergodicity is known as vital importance to the study of quantum many-body systems. A system is ergodic, if a macroscopic variable ultimately evolves in time to its value predicted by the theory of ensemble, and thereafter exhibit small fluctuations around an equilibrium value. Here, we tested the ergodicity analytically on the 1D spin-1/2 anisotropic XY model in a transverse magnetic field by focusing on quench dynamics throughout the  zero-temperature phase space of the system.

It is known the model shows a quantum phase transition from an ordered  FM phase to the PM phase at  a critical point, $h_c$. We let the model experiences quenches in two setups, single quench, and cyclic quench,  with two conditions, choosing the ground state of the system as the initial state, and fixing the system at zero temperature. The Hamiltonian is  integrable and the results are exact. We  considered dynamics of the magnetization along the Z-axis and the spin-spin correlation function along the $X$-axis for both setups. As a consequence we  clearly showed these two order parameters gave the reversely results.
Our results displayed that for single quenches the ergodicity obviously depends on both the initial state and the order parameter. For a certain order parameter, ergodicity can be observed with an ergodic-region for quenches  from a phase, non-correspond to the phase of the order parameter, into itself.  Notably, for quenching from  a phase point corresponding to the order parameter into or very close to the quantum critical point, $h_c=1.0$,  the system behaves ergodic. Otherwise,  for all other single quenches, the non-ergodic behavior of the system appears.  Moreover, we also discussed the inevitable  role of the excited states emerging from the dynamics of the system to create ergodicity. 
Remarkable results obtained for a cyclic quench. In this case, the ergodicity is broken for quenches started from a phase corresponding to the order parameter, independent of the anisotropy parameter, the first quenched point $h_{f_1}$, and the quench time $T$. With a double quench started from a ground-state phase,  non-corresponding to the order parameter, by tuning the elapsing time at a given intermediate phase, one can achieve control over ergodicity in the system. Consequently, our results show for a quantum system at zero-temperature, both ergodicity, and non-ergodicity phenomena can appear dependence of the initial state, the middle quenched point, and also the quench time $T$. It should note we believe these results have a relationship with the conception of heating{\color{blue}\cite{53}}. Therefore, we give a suggestion one can consider the problem from this view.
We hope the presented results provide further evidence for unveiling a relationship between ergodicity, order parameters, and  the non-equilibrium  dynamics of quantum systems.

%\bibliography{123}

\section*{Acknowledgment}
H. Cheraghi is thankful to Dirk Schuricht for providing some notes. This work was done when H. Cheraghi was at Gothenburg University as a visiting researcher. He is very grateful for Henrik Johansson from his so kind of hospitality.
\section*{Author Contributions}
H. Cheraghi proposed the original idea and S. Mahdavifar helped develop and interpret the results. All authors reviewed the manuscript.
\section*{Additional Information}
\textbf{Competing Interests:} The authors declare that they have no competing interests.

\end{document}